\input{aipcheck}

\documentclass[
    ,final            
  ]
  {aipproc}

\layoutstyle{6x9}

\begin{document}

\title{Precise abundance analysis of the outer halo globular cluster M 75}


\author{Nikolay Kacharov}{
  address={Landessternwarte, Zentrum f\"ur Astronomie der Universit\"at Heidelberg, K\"onigstuhl 12, 69117 Heidelberg, Germany}
}

\author{Andreas Koch}{
  address={Landessternwarte, Zentrum f\"ur Astronomie der Universit\"at Heidelberg, K\"onigstuhl 12, 69117 Heidelberg, Germany}
}

\begin{abstract}
Globular clusters (GCs) are the oldest stellar systems in the Milky Way. Long time considered as simple stellar populations, nowadays we recognize their complex star formation history through precise abundance analysis of a variety of chemical elements in individual cluster members. Although we do not necessarily see clues for multiple populations in all GC colour-magnitude diagrams, all GCs present significant spreads and certain anticorrelations between their light and $\alpha$ element abundances. Furthermore, the heavy element abundances in individual stars of the primordial generation and their comparison to halo field stars and dwarf galaxies could provide us with valuable information about the very first stars that could have formed in GCs.
M75 is a unique outer halo (galactocentric distance of $\sim15$ kpc) GC with a peculiar Horizontal Branch morphology. Here we present the first abundance measurements of 16 individual red giants from high resolution spectroscopy. The cluster is metal rich ([Fe/H] $= -1.17 \pm 0.02$), $\alpha$-enhanced, and shows a marginal spread in [Fe/H] of 0.07 dex, typical of most GCs of similar luminosity. The O-Na anticorrelation is clearly visible, showing at least two generations of stars, formed on a short timescale. We also discuss r- and s-process element abundances in the context of the earliest cluster enrichment phases.
\end{abstract}

\maketitle


\subsection{Abundance spreads and multiple populations}

One of the most striking characteristics of all GCs studied to date are the large spreads in their light elements (Fig.~1. Left). Nowadays these spreads and the observed correlations between the p-capture and the $\alpha$-elements are considered as evidence for the existence of at least two generations of stars in the GCs \cite{Gratton+2012}. They can give us clues for the GCs formation and early evolution. The right panel of Fig.~1 shows the Na-O anticorrelation in M 75, as measured from high-resolution (MIKE/Magellan) red giant spectra. Stars with [Na/Fe]~$< 0.1$~dex can be considered the remainder of a Primordial generation that first formed in the GC. These stars are similar to the stars in the galactic halo. In contrast, stars from the second generation show considerably higher Na- and lower O-abundances, but they are indistinct in terms of other chemical elements (Fig.~2).

\begin{figure}
  \includegraphics[height=.21\textheight]{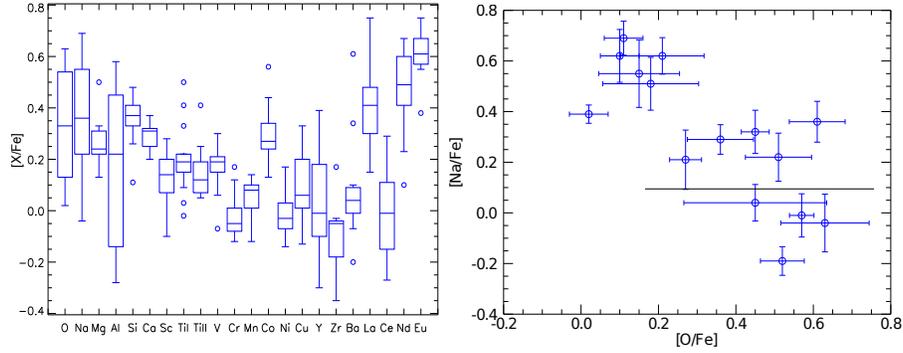}
  \caption{Left: Box-plot of M75's chemical abundance ratios. The horizontal lines indicate the median values, the boxes represent the IQR, and the vertical lines indicate the minimum and maximum values. Outliers are drawn with small circles. Right: O-Na anticorrelation in M 75. The empirical horizontal line separates the two generations \cite{Carretta+2009}.}
\end{figure}

The overall $\alpha$-enhancement of GCs and their small iron spreads speak for a rapid star-formation that ceased before SN Ia began to contribute iron to the cluster environment. But what were the most likely polluters for the 2nd generation? In present day clusters, the fraction of 1st generation stars is small, about $30\%$. The best candidates for enriching the ISM with light elements but without producing significant amounts of $\alpha$ and iron-peak elements are intermediate mass AGB stars or fast rotating massive stars (FRMS). But for both mechanisms to work one has to either invoke a top-heavy IMF for the 1st generation or to assume that the GCs were much more massive and they lost a large fraction of their initial mass. In fact, the similarity in chemistry and ages of the Milky Way halo stars with the P-generation in GCs suggests that the bulk of the halo stars were formed in GCs \cite{Gratton+2012}(and references therein).

\begin{figure}
  \includegraphics[height=.2\textheight]{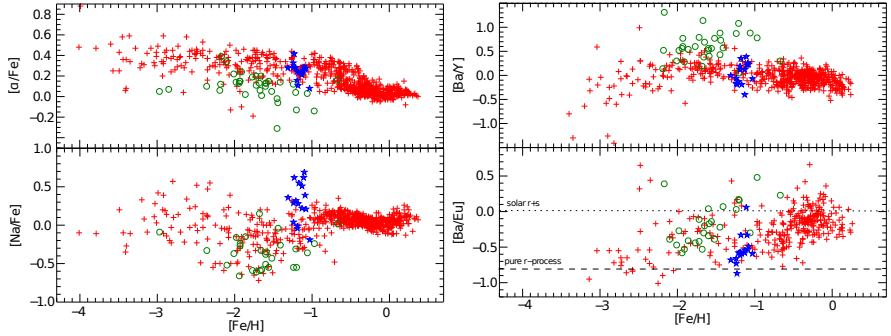}
  \caption{A comparison of the $\alpha$, p-, and n-capture element abundances of the 16 M75 stars (blue asterisks) with Galactic disk and halo stars (red crosses) and individual stars from various dwarf galaxies (green circles) from \cite{Pritzl+2005}}
\end{figure}

\subsection{r- and s-process elements}
 
According to Fig.~3, M 75 is one of the rarer cases of GCs compatible with predominant r-process production of the n-capture elements \cite{Sneden+2000}. An exception are the lighter elements Y and Zr, which are more consistent with scaled solar production. Y and Zr, however, are associated with the weak s-process, which appears in massive (M$\sim20~M_{\odot}$) stars on similar timescales as the r-process production from SNe II. But the [Ba/Y] ratio is similar to that in halo stars (Fig.~3) and thus consistent with a normal IMF \cite{Tolstoy+2009}.

\begin{figure}
  \includegraphics[height=.2\textheight]{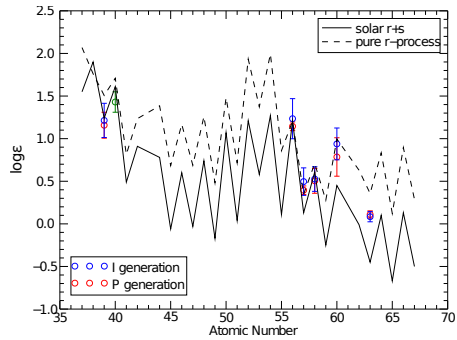}
  \caption{Mean neutron capture elements measurements separately for the P- and I-generations in M 75, normalized to Ba. The lines display the solar and r- and s-process contributions from \cite{Burris+2000}.}
\end{figure}

Fig.~3 shows that there is no difference in the n-capture elements production in the two generations of stars. The lack of evidence of a main s-process in M 75 pulls the scales towards FRMS being the main polluter in light elements for the second generation \cite{Decressin+2007}, while the presence of a few Ba-rich stars indicates that contribution from AGB stars cannot be ruled out. Most probably both processes take place, despite they work on very different timescales.

\subsection{Conclusions}

M 75 hosts at least two chemically distinct generations of stars formed on a short timescale, bringing it in line with other GCs of similar mass and metallicity. Our analysis of the n-capture elements favours a normal IMF for the primordial generation of M 75. It is surprizing that, at [Fe/H] $\sim-1.2$~dex, both generations seem consistent with predominant r-process enrichment. If so, the main polluters for the second generation should be FRMS but the observed large scatter in the [Ba/Eu] indicates that enrichment from AGB stars is also feasible.


\begin{theacknowledgments}
\footnotesize{The authors acknowledge the Deutsche Forschungsgemeinschaft for funding from  Emmy-Noether grant  Ko 4161/1.}
\end{theacknowledgments}



\bibliographystyle{aipproc}   

\bibliography{sample}

\begin{thebibliography}{9}

\bibitem{Gratton+2012}
Gratton, R.~G., Carretta, E., and Bragaglia, A., 2012, A\&ARv, 20, 50


\bibitem{Carretta+2009}
Carretta, E., Bragaglia, A., Gratton, R.~G., et al. 2009, A\&A, 505, 117

\bibitem{Pritzl+2005}
Pritzl, B. J., Venn, K. A., and Irwin, M. J., 2005, AJ, 130, 2140

\bibitem{Sneden+2000}
Sneden, C.,  Johnson, J., Kraft, R.~P., et al. 2000, ApJ, 536, 85

\bibitem{Tolstoy+2009}
Tolstoy, E., Hill, V., and Tosi, M., 2009, ARA\&A, 47, 371

\bibitem{Decressin+2007}
Decressin, T., Meynet, G., Charbonnel, C., et al., 2007, A\&A, 464, 1029

\bibitem{Burris+2000}
Burris, D.~L., Pilachowski, C.~A., Armandroff, at al., 2000, ApJ, 544, 302

\end{thebibliography}

\IfFileExists{\jobname.bbl}{}
 {\typeout{}
  \typeout{******************************************}
  \typeout{** Please run "bibtex \jobname" to optain}
  \typeout{** the bibliography and then re-run LaTeX}
  \typeout{** twice to fix the references!}
  \typeout{******************************************}
  \typeout{}
 }


\end{document}